\documentstyle[epsf,epsfig,wrapfig]{article}
\begin{document}
\vspace{2cm}
\title{ \bf Insufficiency of Available Data on the Behaviour \\
of the Meanlives of Unstable Hadrons with Energy \\[5mm]}
\vspace{1.5cm}
\author{
{\large Yu. Arestov$^{1,2}$, R.M. Santilli$^2$ and V. Solovianov$^1$} \\[3mm]
{\it $^1$ Institute for High Energy Physics, 142284, Protvino, Russia }\\
{\it $^2$ Institute for Basic Research, P. O. Box 1577, Palm Harbor, FL 34682}
}
\date{}
\maketitle
\vspace{0.7cm}
\begin{center}
\parbox {12cm}{
We review the available evidence
according to which physical media alter the Minkowskian spacetime
with consequential alteration of the speed of light;
we point out the apparent emergence of superluminal speeds
within the hyperdense hadrons; and
we point out the lack of conclusive character of the
available related measure, those on
the behaviour of the meanlives of unstable kaons with
energy. \\[0.3cm]
} \\[1cm]
\end{center}
\paragraph*{ Data on local speeds of light.} 

Strictly speaking, the speed of electromagnetic (elm) waves  
is not a "universal constant", but rather a quantity $c = c_o/n$
depending on 
local physical conditions representable via the 
index of refraction n, where $c_o$ is the speed in vacuum. When 
experimentally established, deviations 
from $c_o$ are then rather forceful evidence of deviations from the 
conventional Minkowskian spacetime of the vacuum [1a].

Speeds $c = c_o/n < c_o$ are known in our Newtonian 
environment. Lesser known is the fact that one of the first studies on  
the implications of speeds $c < c_o$ 
were first studied by Lorentz [1b] 
(see the related mention in  
Pauli's book [1c]).

Speeds $c = c_o / n > c_o$ 
have been apparently measured by A. Enders and G. Nimtz [1d]     
in the tunneling of photons between certain guides 
(see review [1e] for additional references and details). Apparent speeds 
$c = c_o/n > c_o$ have also been identified in  
certain astrophysical events [1f-1h] (see also the recent data [1i]). 

Note that the hopes of regaining the exact Minkowskian spacetime  
by reducing light to photons scattering 
among molecules, even though valid as a first approximation, is no
longer 
viable because: 1) the 
reduction to second quantization is questionable for 
elm waves in our atmosphere, say,  
with one meter wavelength; 
2) the reduction does not permit quantitative studies of 
superluminal speeds; and 3) the reduction eliminates the representation
of 
the inhomogeneity and 
anisotropy of physical media, which have apparent,  
experimentally measurable effects (see below). 

Recall that hadrons are not ideal spheres with isolated points in them, 
but rather some of the densest media measured in laboratory until now.
If 
spacetime anomalies are established for media of relatively low density, 
the hypothesis that the Minkowskian spacetime can be {\it exact} within
hadrons 
in its conventional realization has little scientific credibility (see
below 
for the exact character of an axiom-preserving covering spacetime).
Also, 
deviations are expected from the complete mutual penetration of the 
wavepackets of the constituents, thus resulting in the historical open 
legacy of the existence 
of nonlinear, nonlocal and nonpotential effects in the interior of 
hadronic.

One of the first quantitative studies of the above legacy was done by 
D. L. Blokhintsev [2a] in 1964, followed by 
L. B. Redei [2b], D. Y. Kim [2c] and others. 
Note that the exact validity of the 
Minkowskian geometry for the {\it center-of-mass behavior} of a 
hadron in  a particle accelerator is beyond scientific doubts. The
authors of 
Refs. [2a-2c] then argued that 
a possibility for internal anomalies due to 
nonlocal and other effects to manifest themselves in the 
outside is {\it  via deviations from the conventional 
Minkowskian behavior of the meanlives 
of unstable hadrons with the speed} v (or energy E).

Note that the Minkowski metric can be written $\eta = {\rm Diag}. (1, 1, 1,
-c_o^2)$. 
Therefore, {\it any deviation} $\hat{\eta}$ {\it from}  
$\eta$ {\it necessarily 
implies a deviation from} $c_o$, as one can see by 
altering any component of the metric and then using Lorentz transforms.

Along these lines, R. M. Santilli [2d] submitted in 1982 
the hypothesis that {\it contact-nonpotential interactions  
(thus including the strong interactions as per the above legacy)  
 can accelerate 
ordinary (positive) masses at speed bigger than 
the speed of light in vacuun } 
 much along the subsequent   
astrophysical measures [1f-1h].  
The above hypothesis implies that
{\it  photons travel inside the hyperdense hadrons 
at speeds bigger than that in vacuum}. V. de Sabbata and M. Gasperini 
[2e] conducted the first phenomenological verification  within 
the context of the conventional gauge theories  
supporting the hypothesis of Ref. [2d], and actually reaching limit
speeds 
up to $75 c_o$ for superheavy hadrons. 

The above hypothesis is also 
supported by the phenomenological calculations 
conducted by 
H. B. Nielsen and I. Picek [2f] via the spontaneous symmetry breaking  
in the Higgs sector of conventional gauge theories, which have resulted 
in the anomalous Minkowskian metrics (here written in the notation
above)
\begin{eqnarray}
  \pi:~~ \hat {\eta} = {\rm Diag}.[(1 + 1.2\cdot 10^{-3}),~(1 + 1.2\cdot
10^{-3}),~(1 + 1.2\cdot 10^{-3}),~ 
- c_o^2(1 - 3.79\cdot 10^{-3})],\\
 K:~~  \hat{\eta} = {\rm Diag}. [(1 - 2.0\cdot 10^{-4}),~ (1 - 2.0\cdot
10^{-4}),~ (1 - 2.0\cdot 10^{-4}),~ 
- c_o^2(1 + 6.00\cdot 10^{-4})].
\label{(2)}
\end{eqnarray}

As one can see, calculations [2f] 
confirm speeds of photons $c = c_o/n > c_o$  
for the interior of kaons,  
as conjectured in Ref. [2d]. 
Recall that: spacetime anomalies  
are expected to increase with the density; 
all hadrons have approximately the same size;   
and hadrons have densities increasing with mass.  
Therefore, results similar to (2) are expected 
for all hadrons {\it heavier} than kaons,
 as supported by phenomenological studies 
[2e]. 

The first direct experimental measures on the behavior of the 
meanlife of $K_S^o$ with energy, $\tau (E)$, were done by S. H. 
Aronson {\it et al.} [3a] at Fermilab and they 
suggested {\it deviations} from the Minkowskian spacetime 
in the energy range of 
30 to 100 GeV. Subsequent direct measures also for $K_S^o$ were  
done by S. H. Aronson {\it et al.} [3b] 
also at Fermilab, suggesting instead {\it no deviations} of $\tau (E)$ 
from the Minkowskian form in the {\it different} energy range 
of 100 to 400 GeV. 

More recently, a test of
the decay law at short decay times was made by the OPAL group
at LEP [3c]. In the latter experiment  
the ratio of number of events
$Z^0 \rightarrow \tau^{+} \tau^{-}$
with deviations of  $\tau$ from the conventional law to number of
''normal'' events was $(1.1 \pm 1.4 \pm 3.5)\%$.

\paragraph*{ Isominkowskian geometrization of physical media.} 

A geometrization of all deviations from the Minkowskian spacetime 
was submitted by Santilli [4a] in 1983 
under the name of {\it isominkowskian geometry} 
(see [4b] for the latest account) and resulted to be: "axiom-preserving" 
(in the sense that the isominkowskian geometry is isomorphic to 
the conventional one, a property denoted with the prefix "iso");  
"invariant" (in the sense of admitting a 
symmetry isomorphic to the Poincar\'e symmetry [4a-4d];   
and "universal" (in the 
sense of admitting all 
infinitely possible, well behaved, signature-preserving and 
symmetric modifications of the Minkowski metric [4e]). 

Moreover, the isominkowskian geometry has permitted the {\it exact}   
reconstruction of the 
special relativity under 
{\it arbitrary} local speeds 
of light [4f]. Refs. [4] have 
therefore established that, contrary to a popular belief (see, e.g., the 
"Lorentz asymmetry" of Ref. [2f]), the Minkowskian axioms, the Lorentz 
and Poincar\'e symmetry and the special relativity remain {\it exact}
under 
all the above {\it spacetime anomalies}, of course, when properly
formulated.

The isominkowskian geometry is  
essentially characterized by the lifting of the Minkowskian 
metric $\eta \rightarrow \hat{\eta} = \hat T\times \eta$, 
where $\hat T(x, v, E, \mu, \tau, \omega, ...)$ is a 
positive-definite $4\times 4$ matrix with 
 an arbitrary local dependence 
on coordinates x, speeds v, energies E, ensities $\mu$, temperatures
$\tau$, 
frequencies $\omega$, and any other needed variable.  
. Jointly, the 
basic unit of the Minkowski space, $I$ = Diag. (1, 1, 1, 1), is lifted by 
an amount which is the {\it inverse} of the deformation of the metric, 
$I \rightarrow \hat I = 1/\hat T$. The dual lifting 
$\eta \rightarrow \hat{\eta} = \hat T\times \eta$ and $I \rightarrow 
\hat I = 1/\hat T$ then implies the preservation of all 
original spacetime axioms [4] (see Ref. [41-4k] for 
mathematical studies and [4l]  
for physical profiles).

The isominkowskian geometry provides a geometrization of physical media 
at bopth the classical and operator levels [4l]. 
Since $\hat T$ is positive-definite, $\hat{\eta}$ can always 
be diagonalized in the 
form $\hat{\eta}$ = Diag. $(1/n_1^2, 1/n_2^2, 1/n_3^2, -c_o^2/n_4^2)$, 
thus providing 
a geometrization of: the local {\it inhomogeneity} (e.g., via 
a dependence of the n's from the density); 
the local {\it anisotropy} (e.g., via  
a differentiation between the space and time n's); as well as 
{\it arbitrary local speeds of elm waves} (via the expression  
$c(x, \mu, \omega, ...) = c_o/n_4(x, \mu, \omega, ...)$ 
first proposed in [4a]).
 
The isotopic  
behavior of the meanlife with speed (or energy) for isotropic space with 
$n_1 = n_2 = n_3 = n_s(x, \mu, \omega, ...)$ (yet with 
general spacetime anisotropy $n_s \not = n_4$)  is given by [4a-4c]
\begin{equation}
 \hat{\tau} = \tau_{o}\hat\gamma, ~\hat\gamma = (1-\hat\beta^2)^{-1/2}, 
~\hat\beta
 =  (v/n_s) / (c_{o}/n_4),  
\label{(3)}
\end{equation}

\noindent and includes all existing 
or otherwise possible laws [2] via 
different power series expansions 
in terms of different parameters with different truncations [4e]. This 
eliminates the ambiguity of individually testing the several different
laws 
of Refs. [2]. 

Note that, when a hadron is studied  
from the outside, one evidently can only use the {\it average of the 
n-quantities to constants}, called ''characteristic constants'' of the 
medium considered. Note also that a possible anysotropy of the medium 
implies a deviation from the conventional Doppler shift studied by
Mignani [5a] 
and others which will be studied elsewhere as a possible 
complement to measures [2,3]. Note finally that the latter anomalies are 
eliminated by the reduction of  
of light to photons {\it  moving in vacuum} and scattering among
molecules.

Isotopic law (3) was applied by F. Cardone, R. Mignani and R. M.
Santilli [5b]  
to the experimental data of Refs. [3a,3b] 
resulting in the single fit of both experiments,  
\begin{equation}
 1/n_1^2 = 1/n_2^2 = 1/n_3^2 = 0.909080 \pm 0.004,~~ 1/n_4^2 = 1.003 
\pm 0.002.
\label{(4)}
\end{equation}

Therefore, even under the assumption of the correct character of 
measures [3b], they do not establish the validity of the 
Minkowskian geometry inside hadrons because of the above 
isominkowskian fit. Note also that fit (4) confirms the 
superluminal character of the propagation 
of light within the hyperdense hadronic media, a property that 
appears to be confirmed by other studies (see the outline in [4b]). 
We should finally mention that nonlinear and nonlocal effects 
at short distances have been recently studied in Refs. [5d,5e,4l].

\paragraph*{ An alternative data elaboration.}

In this note we focus the attention on the range-energy selection rule
which can be applied to re-elaborate the initial data on $K_S$
decays from the experiment [3b].
 By taking into account the results
as they were done, we performed Monte Carlo simulations
of the main features of experiment [3b] and made our own fits for
$K_S^o$.
Our conclusions and recommendations are the following: \\
\indent 1) We agree 
  that the parameters in the full formula $dN/dt$ for the proper time
evolution are strongly
correlated. This may cause a generally non-relevant regular dependence
of the parameters on entities which are not present in the formula, such
as
number of runs, energy, etc., apart from the systematic uncertainties.
Therefore, the above dependence
may shadow the weak energy dependence we are interested in,
as can be seen from the large values of the correlation elements.

   2) The authors of Ref.~[3b] solved the problem of non-correlated fit
by selecting the $K_S^o$ momenta greater than 100 GeV/c. By means of
that
energy cut, they obtained the data sample in which the CP violating
terms
contribute up to 1.6\%.
However, it seems unrealistic to look for the deviations from the
Minkowskian decay law of the order of some percent.
More realistic is to test the decay law on the level of $10^{-3}$,
as suggested by studies [2].
 In fact,
the assumption of 1.6\% contribution from PC violation
in the data elaboration of Ref. [3b]
implies looking for the energy dependence of $\tau_s$
at the level $k\cdot 10^{-2}$, thus rendering meaningless to look
for more realistic deviations of the order of $10^{-3}$ and smaller.

3) we propose to suppress
the CP violating terms significantly using selection rule for the ratio
$R/E$,
where $R$ and $E$ are $K_S^o$ range and energy.
In the  experiment, $R/E$ ranges from 2.3 to 36.1 cm/GeV. The $R/E$
interval
should be selected to make the contribution of the CP violating terms
less
than a desirable value, say $k\cdot 10^{-3}$. An effective ($R,E$)
plot can then be
calculated via Monte Carlo methods applied to the real decay volume.\\
\vspace{-3mm}
\begin{wrapfigure}{r}{8.0cm}
%
\epsfig{figure=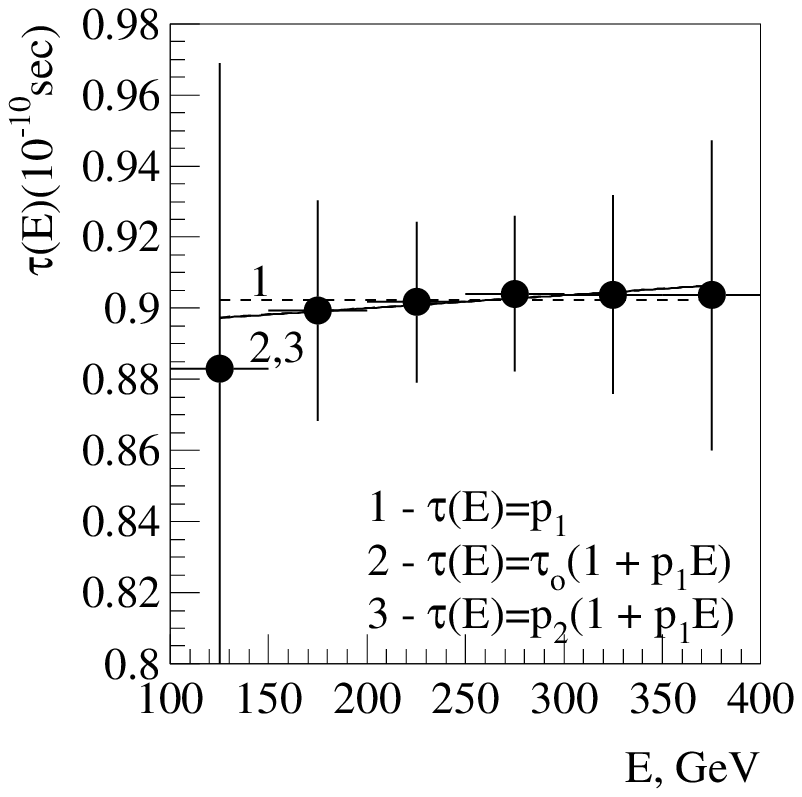,bbllx=185pt,bblly=280pt,bburx=400pt,%
bbury=520pt,width=7.5cm,height=8cm}
%
{ Fig. 1. 
Comparison of the various fitting functions (curves 1,2 and 3)
applied to the simulated lifetime $\tau$(E) dependence in ref.~3b
under the energy-range selection rule (see the text).
}
\end{wrapfigure}

   The price we pay for more accurate data handling due to
the range-energy selection rule will be {\it lower statistics.}
In fact, under the above new assumptions,
60-70\%  events will be rejected, i.e.,
only 63K - 84K events of the total 220K will be useful.
   Apart from the loss of a major part of the data, 1/3 of the
decay volume in the experiment turns out to be also useless.
   The large inefficiency of the experiment occurred because it had not
been optimized for the problem. Basicly, the experimental design and
data
selection rules followed that of conventional $K_S,\, K_L$ studies.

   We illustrate the above arguments with two fits shown in Fig. 1. 220
000
$K_S$ decays at six energy values (from 125 to 375 GeV) were generated
in
the decay volume with the ranges from 9.3 m to 25.3 m.
The energy dependence of the lifetime was assumed in the form
$\tau(E)=\tau_S(1+\epsilon E)$ with $\tau_S$~=~0.8927, the world average
of
the mean lifetime, and $\epsilon$~=~4$\cdot 10^{-5}$.
After applying the
range-energy selection rule, a sample of 64K events was chosen for which
the contribution of the CP violating terms was less then 0.008. Namely
we
deal with the following distribution for the proper lifetime:
\begin{eqnarray}
\frac{dN}{dx} = N\lbrace \exp{(-x)}~+~{\rm CPV} \rbrace,
\end{eqnarray}
where $N$ is a normalization constant, $x=t/\tau(E)$ and CP violating
terms are equal to
\begin{eqnarray}
\nonumber
 {\rm CPV}~=~\mid \eta_{+-} \mid ^2 {\rm exp}(-xy)~+~2D\mid \eta_{+-}\mid  
~{\rm cos}(\Delta m~t - \phi_{+-}){\rm exp}(-x(1+y)/2)
\end{eqnarray}
where $y$ stands for $\tau_S (E)/\tau_L$.

The values of other parameters
are taken as the world average values. These are
 $\mid \eta_{+-} \mid$~=~2.284$\cdot 10^{-3}$, the magnitude of the
CP-nonconservation parameter in $ K_{L}^{o}\to \pi^+ \pi^-$ decay,
$\phi_{+-}$=43.7$^o$, and $\Delta m=0.5333\cdot 10^{10}$ 
$\hbar sec^{-1}$
is the mass difference of $K_L^o - K_S^o$. The dilution factor $D$ is
defined
as the ratio $(N-\bar{N})/(N+\bar{N})$ where $N~(\bar{N})$ is the number
of
$K^o ~(\bar {K^o})$ produced by the proton beam on the target. We
accepted
the value $D$=0.75.

   In Fig.~1 the sequence of the  mean proper lifetimes is plotted
versus $E$,
$K_S^o$ laboratory energies. The dependence was obtained by simulations
of $K_S^o$  decays in the experimental volume under
the conditions described above. The figure presents also results of three
fitting procedures: \\
a) one-parameter fit by a constant function
$\tau (E) =c$ with $c$=0.90$\pm$0.01 and $\chi^2$/ndf~=~0.7/5 (dashed line 1);
\\
b) one-parameter fit by  the energy-dependent formula 
of the type $\tau (E)=0.8927 (1~+~p_1 E)$ 
with the obtained value of the parameter 
$p_1$~=~(4~$\pm$~5)$\cdot 10^{-5}$ and $\chi^2$/ndf~=~0.38/5 (solid line 2);\\
c) two-parameter fit to the formula of Ref.~[3b],
$\tau (E)~=~p_2 (1~+~p_1 E) $. In this case, the value of the crucial parameter
$p_1$ is equal to (4~$\pm$~23)$\cdot 10^{-5}$ with 
$\chi^2$/ndf~=~0.38/4 (dotted line 3 which coinsides practically with solid 
line 2).

   There is a difference in interpretation of parameters in the two
fitting
formulae with the energy dependence. The parameter $p_2$ in the fit from
the cited paper was interpreted
as the zero-energy mean value of the proper  lifetime. We think that it
is
difficult to extrapolate  the fitting formulae from the energy interval
100-400 GeV to zero. Instead, we try to find the energy dependence in
the limited
energy interval by fit starting from a definite point. This
difference in interpretation is important because, in general, various
approaches in fitting procedures may lead to crucially
different numerical results.


   Thus, in the selected amount of the events,  both fits dig up well
the mean
value of the hidden parameter $\epsilon$
determining the energy dependence in the simulated $K_S^o$ decays,
however
the error bars differ strongly. Though both results for fitting values
of
$p_1$
are still  insignificant statistically even in the selected sample of
events,
the 100\% error bar in our fit being rather promising. It opens the door
for
new manipulations with the selection procedure aiming to improve the
result. So we encourage the re-elaboration of the original data
of ref.~[3b] under the modified selection rules to obtain possible
hopeful
estimations of $\tau (E)$ instead of previous hopeless ones.

We finally note that no firm spacetime anomalies can be 
established via the above re-elaboration for PC violating contributions 
smaller than $1.6\%$ because said anomalies are visually within the 
errorbars (Fig. 1) due to insufficient statistuics and other reasons. 
Corresponding deviations cannot be considered for PC violating
contribution 
larger than $1.6\%$ because the latter are experimentally known to be 
excluded for the energy range of measures [3b]. Despite that, the  
analysis of this note 
establishes the insufficiencies of both measures [3a,3b] and the need 
for novel more accurate measures.


\end{document}